\documentclass[referee]{raa}            
\usepackage{graphicx,times}             
\usepackage{lineno}
\usepackage[caption=false]{subfig}
\usepackage{multirow}
\usepackage{graphicx}
\usepackage{booktabs}
\usepackage{threeparttable}
\usepackage{natbib}
\usepackage{rotating}
\usepackage{amsmath,amssymb}
\usepackage[T1]{fontenc}
\begin{document}

   \title{Lobe-dominated $\gamma$-ray Emission of Compact Symmetric Objects}

   \volnopage{Vol.0 (200x) No.0, 000--000}      
   \setcounter{page}{1}          

   \author{Ying-Ying Gan
      \inst{1}
   \and Hai-Ming Zhang
      \inst{2}
   \and Xing Yang
      \inst{3}
   \and Ying Gu
      \inst{3}
   \and Jin Zhang
      \inst{1\dag}
   }

   \institute{School of Physics, Beijing Institute of Technology, Beijing 100081, People's Republic of China; j.zhang@bit.edu.cn
     \and
         School of Astronomy and Space Science, Nanjing University, Nanjing 210023, People's Republic of China
     \and
         Guangxi Key Laboratory for Relativistic Astrophysics, School of Physical Science and Technology, Guangxi University, Nanning 530004, People's Republic of China
   }

   \date{Received~~201X month day; accepted~~201X~~month day}

\abstract{The $\gamma$-ray emitting compact symmetric objects (CSOs) PKS 1718--649, NGC 3894, and TXS 0128+554 are lobe-dominated in the radio emission. In order to investigate their $\gamma$-ray radiation properties, we analyze the $\sim$14-yr Fermi/LAT observation data of the three CSOs. They all show the low luminosity ($10^{41}-10^{43}$ erg s$^{-1}$) and no significant variability in the $\gamma$-ray band. Their $\gamma$-ray average spectra can be well fitted by a power-law function. These properties of $\gamma$-rays are clearly different from the $\gamma$-ray emitting CSOs CTD 135 and PKS 1413+135, for which the $\gamma$-rays are produced by a restarted aligned jet. In the $L_{\gamma}-\Gamma_{\gamma}$ plane, the three CSOs are also located at the region occupied by radio galaxies (RGs) while CTD 135 and PKS 1413+135 display the similar feature to blazars. Together with the similar radio emission property to $\gamma$-ray emitting RGs Cen A and Fornax A, we speculate that the $\gamma$-rays of the three CSOs stem from their extended mini-lobes. The broadband spectral energy distributions of the three CSOs can be well explained by the two-zone leptonic model, where their $\gamma$-rays are produced by the inverse Compton process of the relativistic electrons in extended region. By extrapolating the observed Fermi/LAT spectra to the very high energy band, we find that TXS 0128+554 among the three CSOs may be detected by the Cherenkov Telescope Array in future.
\keywords{galaxies: active---galaxies: jets---gamma rays: galaxies---radiation mechanisms: non-thermal}
}

   \authorrunning{Ying-Ying Gan et al. }            
   \titlerunning{$\gamma$-ray Emitting CSOs }  

   \maketitle
%
%
\section{Introduction}           
\label{sect:intro}

Compact symmetric objects (CSOs), a sub-class of active galactic nuclei (AGNs), are distinguished by a symmetric and compact radio structure with the projection size $\leq$1 kiloparsec (kpc). Their symmetric radio structure is thought to be due to a misaligned jet with no or weak relativistic effect (\citealt{1980ApJ...236...89P, 1994ApJ...432L..87W, 1996ApJ...460..634R}). And thus CSOs look like the mini-version of off-axially observed radio galaxies (RGs). However, some CSOs show obvious flux variation at radio band, even significant variability and high luminosity in $\gamma$-ray band, indicating the strong Doppler boosting effect of a relativistic jet (\citealt{2021RAA....21..201G, 2022ApJ...939...78G}). Recently, \cite{2023arXiv230311357K} reported that in addition to the compact symmetric structure, no significant flux variation and superluminal motion should be taken as a criteria for the CSO selection, that is to say no relativistic effect. 

It is well known that blazars are main one of the extragalactic $\gamma$-ray emitters since a relativistic jet dominates their multi-band radiations, and their $\gamma$-rays are from an aligned pc/sub-pc jet. Generally, it is believed that the $\gamma$-rays of RGs are also generated from their pc/sub-pc jets (e.g., \citealt{2012ApJ...751L...3G, 2012ApJ...746..151A, 2012ApJ...746..141A, 2015ApJ...808..162C, 2015ApJ...799L..18T, 2017RAA....17...90X}). Nevertheless, the detections of $\gamma$-rays from the large-scale lobes of RGs Cen A and Fornax A by the Fermi/LAT indicate the existence of high-energy particles in these extended regions (\citealt{2010Sci...328..725A, 2016ApJ...826....1A}). Until now, five CSOs have been detected in the $\gamma$-ray band, including PKS 1718--649 (\citealt{2016ApJ...821L..31M}), NGC 3894 (\citealt{2020A&A...635A.185P}), TXS 0128+554 (\citealt{2020ApJ...899..141L}), CTD 135 (\citealt{2021RAA....21..201G}), and PKS 1413+135 (\citealt{2021MNRAS.507.4564P, 2022ApJ...939...78G}). The high luminosity and remarkable variability of CTD 135 and PKS 1413+135 observed by the Fermi/LAT, similar to blazars, demonstrate that their $\gamma$-rays are dominated by the radiation of a recently restarted aligned core-jet (\citealt{2021RAA....21..201G, 2022ApJ...939...78G}). Theoretically, the mini-lobes of CSOs can also produce the strong $\gamma$-ray emission via inverse-Compton (IC) processes (\citealt{2008ApJ...680..911S, 2014ApJ...780..165M}), or by thermal bremsstrahlung radiation (\citealt{2007MNRAS.376.1630K, 2009MNRAS.395L..43K}), even through the hadronic processes (\citealt{2011MNRAS.412L..20K}), similar to the large-scale lobes of Cen A and Fornax A (\citealt{2010Sci...328..725A, 2015MNRAS.446.3478M}).

The core-dominance parameter, which is defined as the luminosity ratio ($R_{\rm CE}$) of core to extended region, $R_{\rm CE}>1$ for core-dominated and $R_{\rm CE}<1$ for lobe-dominated, is often used as an indicator of beaming effect (e.g., \citealt{1995PASP..107..803U}). We note that Cen A and Fornax A have the low values of $R_{\rm CE}$ at radio band comparing with other $\gamma$-ray emitting RGs. The $R_{\rm CE}$ values at 8 GHz\footnote{The data are taken from the Radio Fundamental Catalog (RFC), http://astrogeo.org/rfc/.} are 0.48, 0.36 and 0.71 for PKS 1718--649, NGC 3894, and TXS 0128+554, respectively. Especially for PKS 1718--649, even no clear radio core is resolved (\citealt{2019A&A...627A.148A}), resembling a mini-version of Fornax A (very weak radio core, \citealt{2015MNRAS.446.3478M}). Whereas, $R_{\rm CE}$ is $\sim$1 for CTD 135 and 1.42 for PKS 1413+135. It seems to be coincident with that PKS 1413+135 and CTD 135 are reclassified as blazars rather than CSOs (\citealt{2021ApJ...907...61R, 2022Symm...14..321F}). Different from CTD 135 and PKS 1413+135, PKS 1718--649, NGC 3894, and TXS 0128+554 definitely satisfy the CSO criteria asked by \cite{2023arXiv230311357K}. In this paper, we intend to explore the $\gamma$-ray radiation properties of the three lobe-dominated CSOs by completely analyzing their Fermi/LAT observation data. Throughout, $H_0=71$ km s$^{-1}$ Mpc$^{-1}$, $\Omega_{\rm m}=0.27$, and $\Omega_{\Lambda}=0.73$ are adopted in this paper.

\section{lobe-dominated gamma-ray Emitting CSOs}

\emph{PKS 1718--649}, also named NGC 6328 and located in a compact gas environment with redshift of $z=0.014$ (\citealt{1977MNRAS.179...89F}). It is also classified as a GHz-peak spectrum (GPS) source due to a turnover frequency of $\sim$3 GHz at radio spectrum (\citealt{1997AJ....113.2025T}). Its morphology at multiple radio bands is characterized as two sub-structures from southeast to northwest (\citealt{1997AJ....113.2025T, 2002ApJS..141..311T}), and the total projection size at 8.4 GHz and 22.3 GHz is only 2.5 pc (\citealt{2019A&A...627A.148A}). No clear radio core is resolved at multiwavelength observations. However, the spectral index map from 8.4 GHz to 22.3 GHz (Figure 6 in \citealt{2019A&A...627A.148A}) reveals that the region between two sub-structures displays a highly inverted spectrum and may be corresponding to the core, which cannot be observed at these radio frequencies due to synchrotron-self-absorption (\citealt{2019A&A...627A.148A}). The estimated kinematic age is 70$\pm$30 yr via the apparent separation speed of $0.13 \pm 0.06~c$ between two mini-lobes (\citealt{2019A&A...627A.148A}, see also \citealt{2009AN....330..193G}).  

\emph{NGC 3894}, located in an elliptical galaxy with redshift of $z=0.01075$ (\citealt{2014ApJS..210....9B}). Its radio spectrum peaks at a frequency near 5 GHz (\citealt{1998ApJ...498..619T}). The high-resolution Very Long Baseline Array (VLBA) image at 1.4 GHz shows its radio core, the two-sided jets, and a southeast radio lobe with the total linear size of $\sim$50 mas (\citealt{1998ApJ...502L..23P}). The Very Long Baseline Interferometry (VLBI) observations at 5 GHz covering 15 yr demonstrate that the two-sided pc-scale jets emerge from the radio core with a mildly relativistic speed of $\sim 0.3~c$ (\citealt{1998ApJ...498..619T}). \cite{2020A&A...635A.185P} reported that its two-sided jets expand to southeast and northwest directions with an apparent separation velocity of $0.202\pm0.005~c$ between 1995 and 2017 and then estimated its kinematic age to be 59$\pm$5 yr.

\emph{TXS 0128+554}, hosted in an elliptical galaxy with redshift of $z=0.0365$ (\citealt{2012ApJS..199...26H}). The turnover frequency of its radio spectrum is less than 1 GHz (\citealt{2020ApJ...899..141L}). Its pc-scale radio structure is consisted of an unresolved bright core and east-west side cambered lobes, and the overall projected extent of its lobes is similar in the multiple-radio-frequency images (\citealt{2020ApJ...899..141L}). With the observations at 2.3 GHz and 5 GHz, the faint steep-spectrum emission is revealed for the regions at 8.8 pc (eastern lobe) and 7.8 pc (western lobe) from the core (\citealt{2020ApJ...899..141L}). \cite{2020ApJ...899..141L} reported an apparent advance speed of 0.32$\pm0.07~c$ of its radio sub-structure and estimated its kinematic age of 82$\pm$17 yr under the assumption of a constant advance speed in the western lobe.

\section{Fermi/LAT Data Analysis and Results}

\subsection{Data Analysis}

We use the public software \textit{fermitools}\footnote{https://fermi.gsfc.nasa.gov/ssc/data/analysis/software/} (ver. 2.0.8, \citealt{2019ascl.soft05011F}) with the binned likelihood analysis method to analyze the Fermi/LAT Pass 8 data of the three CSOs. The data covering from 2008 August 4 to 2022 May 1 (MJD 54682--59700) are extracted from the Fermi Science Support Center. We set the region of interest as a circle of radius $15^{\circ}$ centered on the radio/optical positions of the three CSOs. The photon events in the energy range within 0.1--300 GeV are considered. We filter the background $\gamma$-ray contamination from the Earth limb by setting the maximum zenith angle as $90^{\circ}$. The good quality photon events are selected by a standard data quality selection criteria ``(DATA\_QUAL$>$0)$\&\&$(LAT\_CONFIG==1)". The data are binned with a pixel size of $0.2^{\circ}$ and 25 logarithmic energy bins. The instrument response function P8R3\_SOURCE\_V3 is used in the data analysis.

The Fermi/LAT 12-yr Source Catalog (4FGL-DR3, \citealt{2022ApJS..260...53A}), the diffuse Galactic interstellar emission (gll\_iem\_v07.fits), and the isotropic emission (iso\_P8R3\_SOURCE\_V3\_v1.txt) are included as background models. In the data analysis, we use the maximum test statistic (TS) to quantify the significance of the $\gamma$-ray detection. TS is the logarithmic ratio of the likelihood of a model with the point source to that of a model without the point source, i.e., TS = $2{\rm log}(\frac{\mathcal{L}_{\rm src}}{\mathcal{L}_{\rm null}})$ (\citealt{1996ApJ...461..396M}). A threshold of TS $>25$ is adopted to recognize a point source from the background. By subtracting all the background models, we firstly generate the $3^{\circ} \times 3^{\circ}$ residual TS maps to investigate whether there are new $\gamma$-ray sources not included in the 4FGL-DR3. No excess $\gamma$-ray signal with TS $>25$ is found in the residual TS maps, indicating that no new additional $\gamma$-ray source is detected. We then produce the $2^{\circ} \times 2^{\circ}$ TS maps of the three 4FGL point sources, 4FGL J1724.2--6501, 4FGL J1149.0+5924, and 4FGL J0131.2+5547. Using the \emph{gtfindsrc} tool, we also obtain the best-fit positions of the three 4FGL point sources, as shown in Figure \ref{TSmap}. Information on the three CSOs and the three associated 4FGL point sources is also given in Table 1.

During the spectral analysis, the normalization of the isotropic emission and the diffuse Galactic interstellar emission, and the spectral parameters of the $\gamma$-ray sources listed in 4FGL-DR3 within the range of $8^{\circ}$ remain free, and other parameters are kept fixed. A spectrum should be significantly curved if TS$_{\rm curv}>9$ (corresponding to a $3\sigma$ confidence level, \citealt{2022ApJS..260...53A}), where TS$_{\rm curv}=2{\rm log}(\frac{\mathcal{L}_{\rm LP}}{\mathcal{L}_{\rm PL}})$, $\mathcal{L}_{\rm LP}$ and $\mathcal{L}_{\rm PL}$ are the hypothesis likelihoods of LogParabole model and power-law model testing, respectively. The TS$_{\rm curv}$ values of the three CSOs are much less than 9, therefore, the power-law spectral model is taken into account to fit their photon spectra in our analysis. The average spectra of the three CSOs with the power-law fits are presented in Figure \ref{Spectra-LC}, and the derived parameters are listed in Table 1. We also generate their long-term light curves with the power-law spectral fits, where the equal time interval of 180 days is adopted, as displayed in Figure \ref{Spectra-LC}.

We also calculate the corresponding variability index TS$_{\rm var}$, a common method to estimate the $\gamma$-ray variability (\citealt{2022ApJS..260...53A}). The definition of TS$_{\rm var}$ is
\begin{equation}
\text{TS}_\text{{var}} = 2\sum_{i=0}^{N} [\text{log}(\mathcal{L}_{i}(F_{i}))-\text{log}(\mathcal{L}_{i}(F_{\rm glob}))],
\end{equation}
where $F_{i}$ is the fitting flux for bin $i$, $\mathcal{L}_{i}(F_{i})$ is the likelihood corresponding to bin $i$, $F_{\rm glob}$ is the best-fit flux for the global time by assuming a constant flux, and $N$ is the number of time bins. We keep only the normalization of the target source free to vary and fix other parameters to the best-fit values during calculations. TS$_{\rm var}=53.0$ corresponds to a $3\sigma$ confidence level in the $\chi^2_{N-1}$(TS$_{\rm var}$) distribution with $N-1=26$ degrees of freedom, where $N=27$ is the number of time bins. The $\gamma$-ray emission of a source is probably variable when TS$_{\rm var}>53.0$. The obtained TS$_{\rm var}$ values are listed in Table 1, and the corresponding confidence levels of variability are also given. 

\subsection{Analysis Results}

\begin{figure}
 \centering
   \includegraphics[angle=0,scale=0.21]{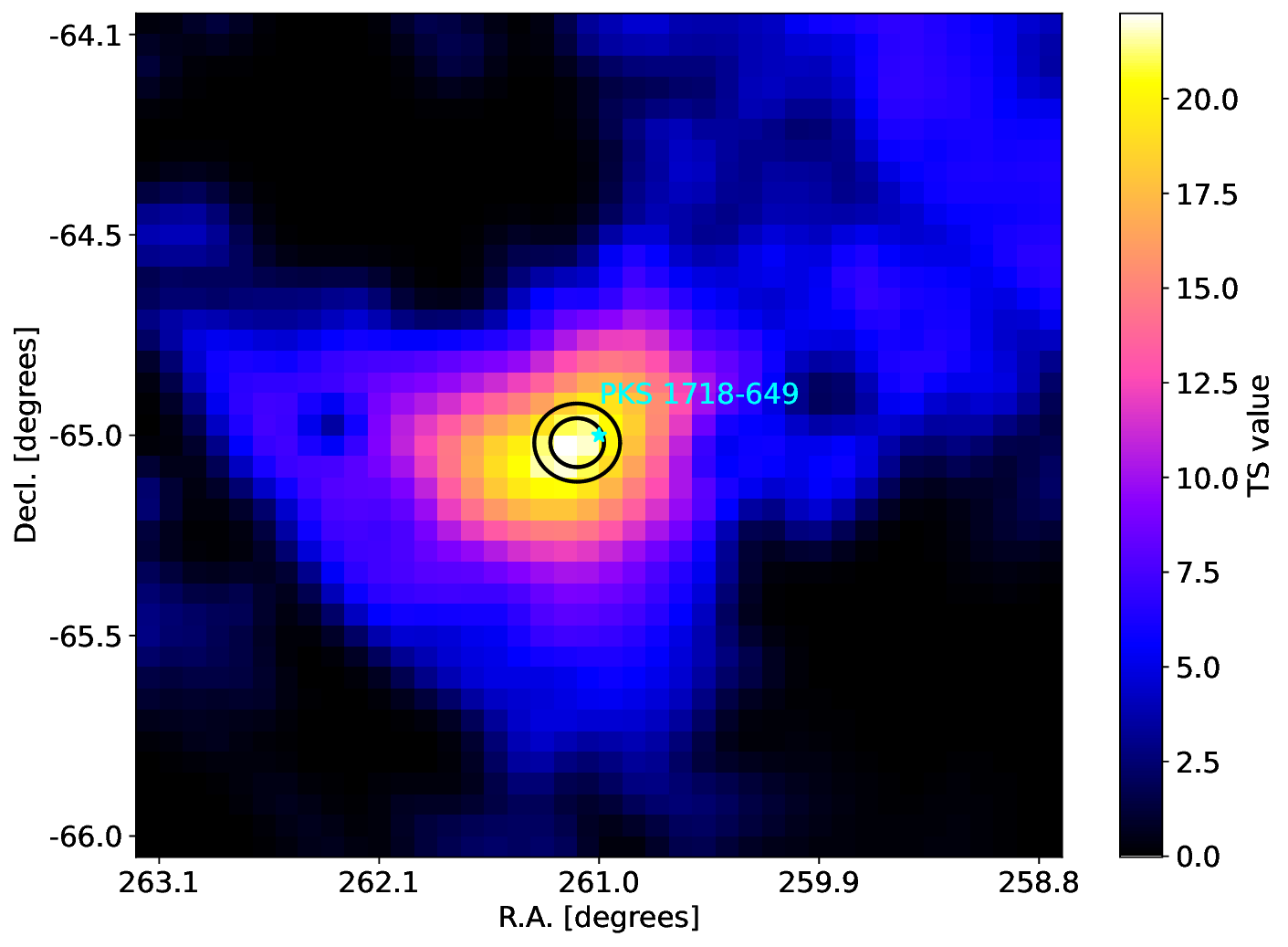}
   \includegraphics[angle=0,scale=0.21]{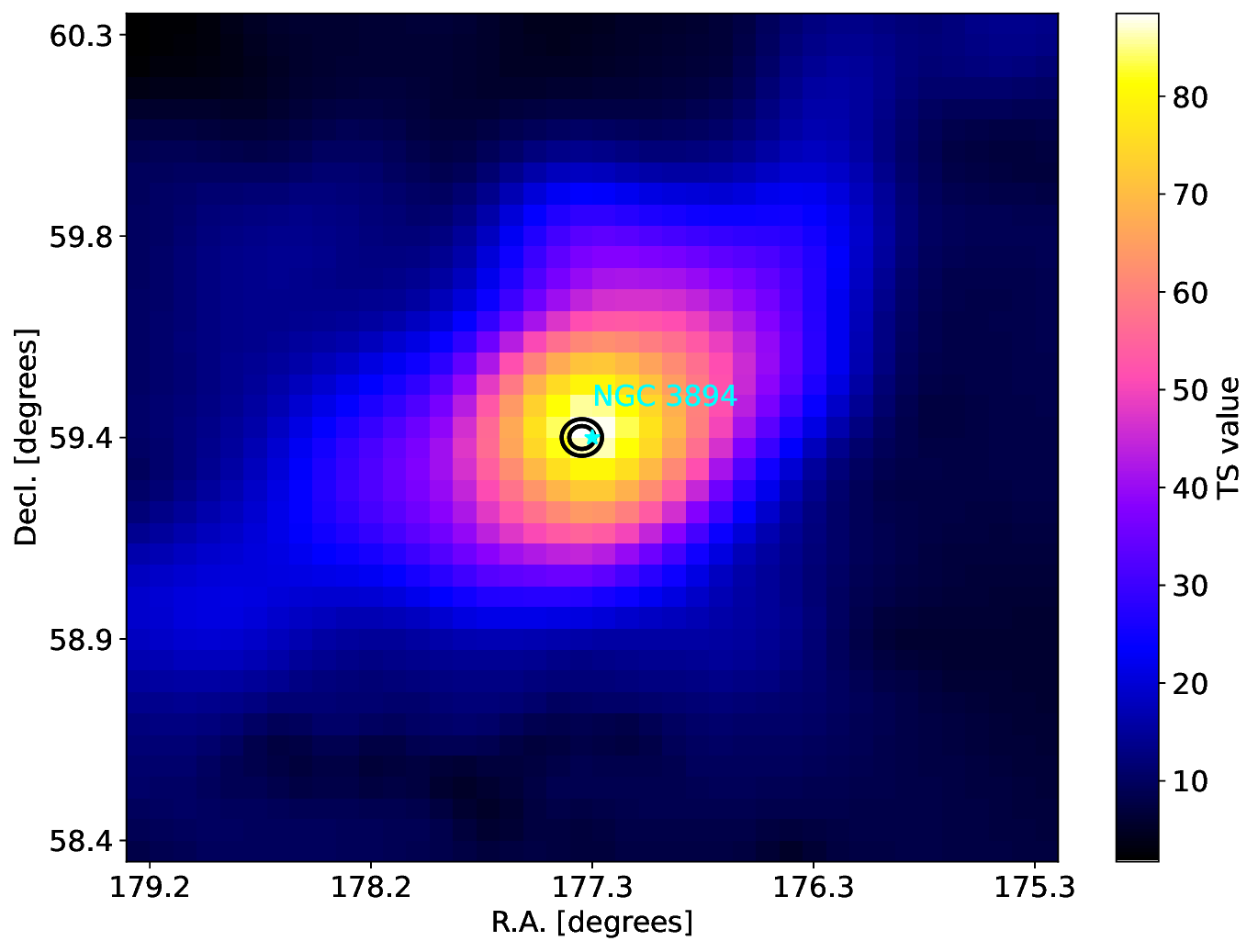}
   \includegraphics[angle=0,scale=0.21]{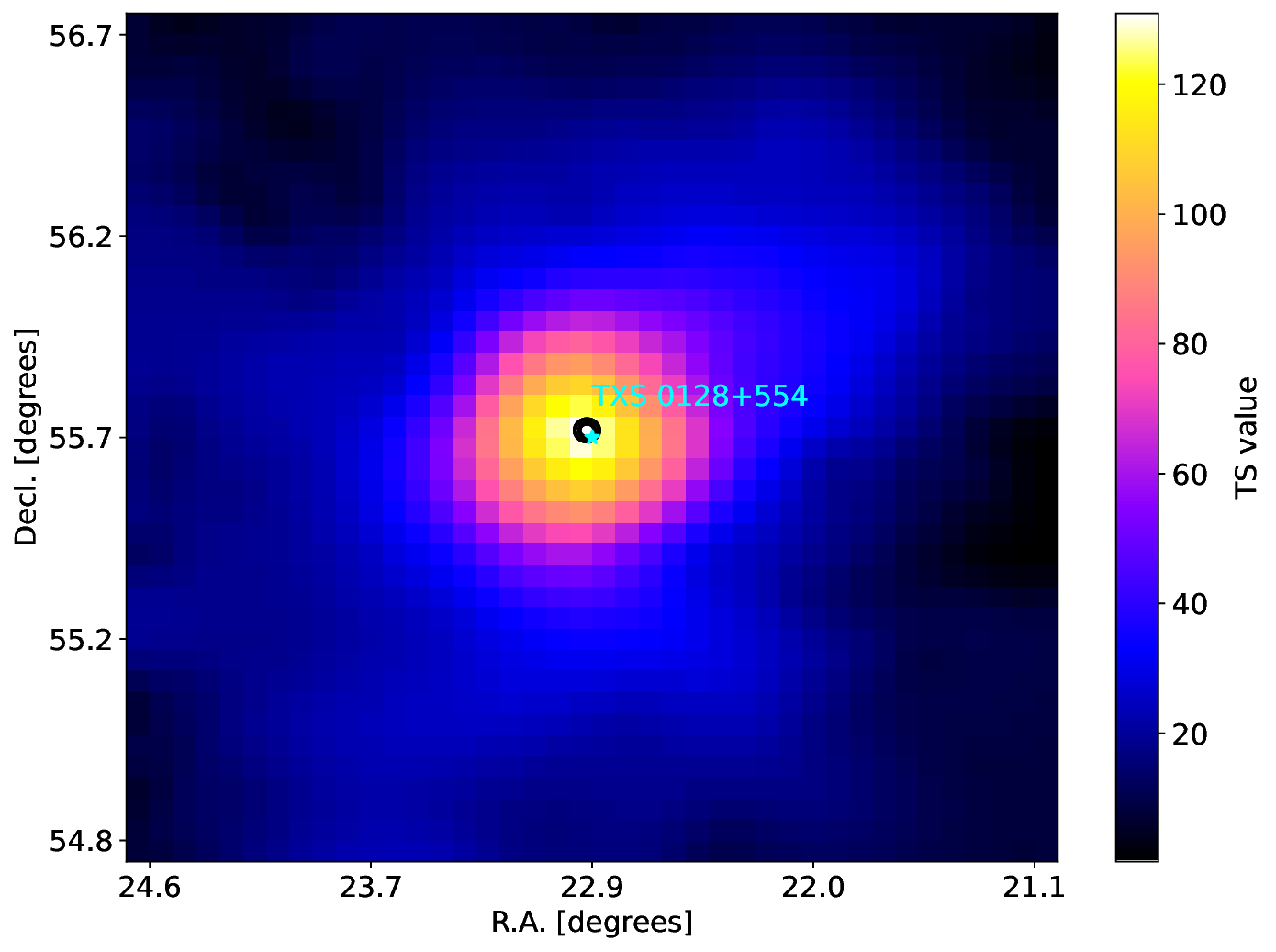}
\caption{$2^{\circ} \times 2^{\circ}$ TS maps of the three 4FGL-DR3 point sources with the $68\%$ and $95\%$ containments (black contours) of the best-fit positions. The cyan solid stars represent the radio/optical positions of the three CSOs. The maps are created with a pixel size of $0.05^{\circ}$ and are smoothed.}
\label{TSmap}
\end{figure}

\begin{figure}
 \centering
   \includegraphics[angle=0,scale=0.27]{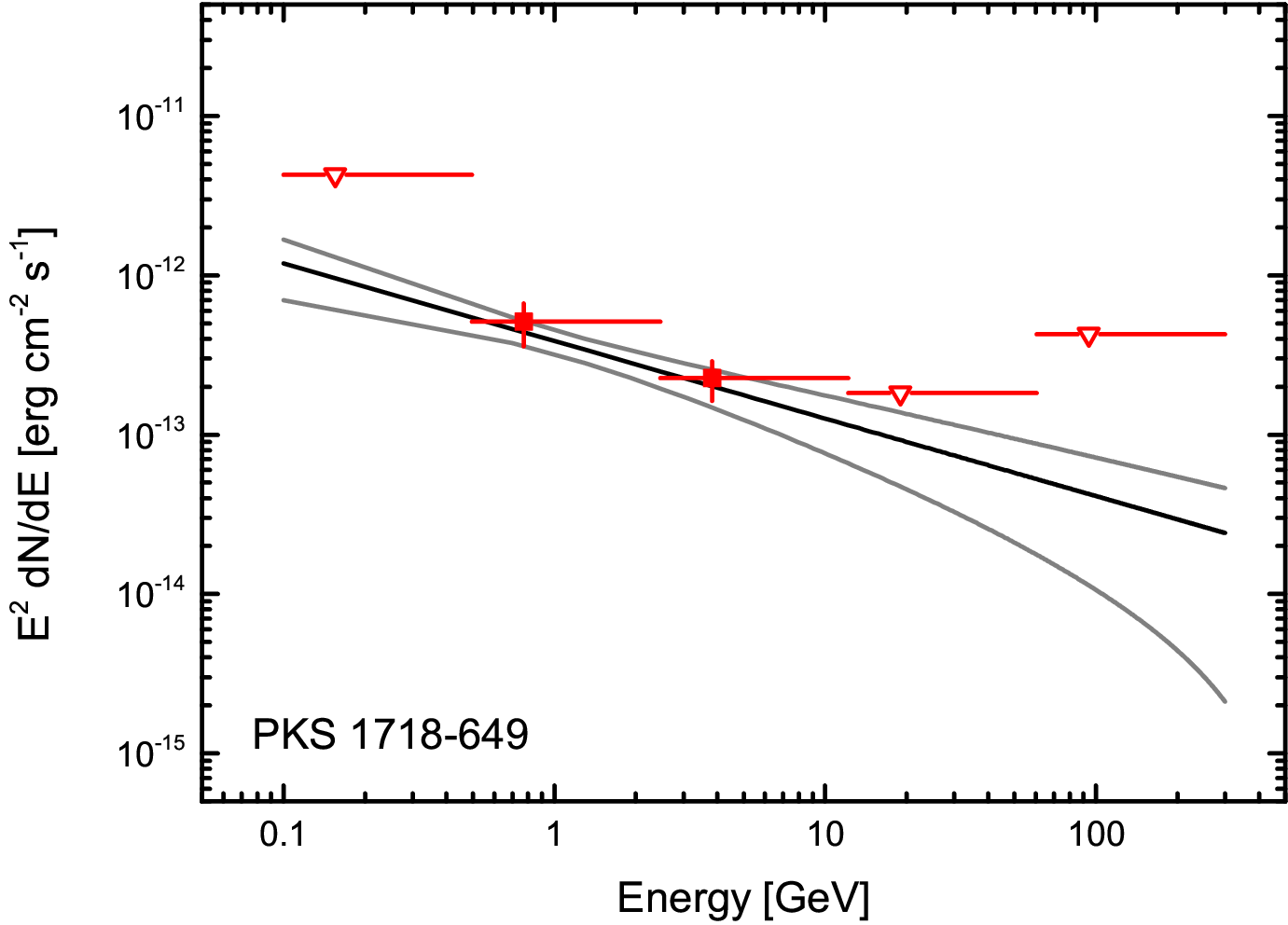}
   \includegraphics[angle=0,scale=0.3]{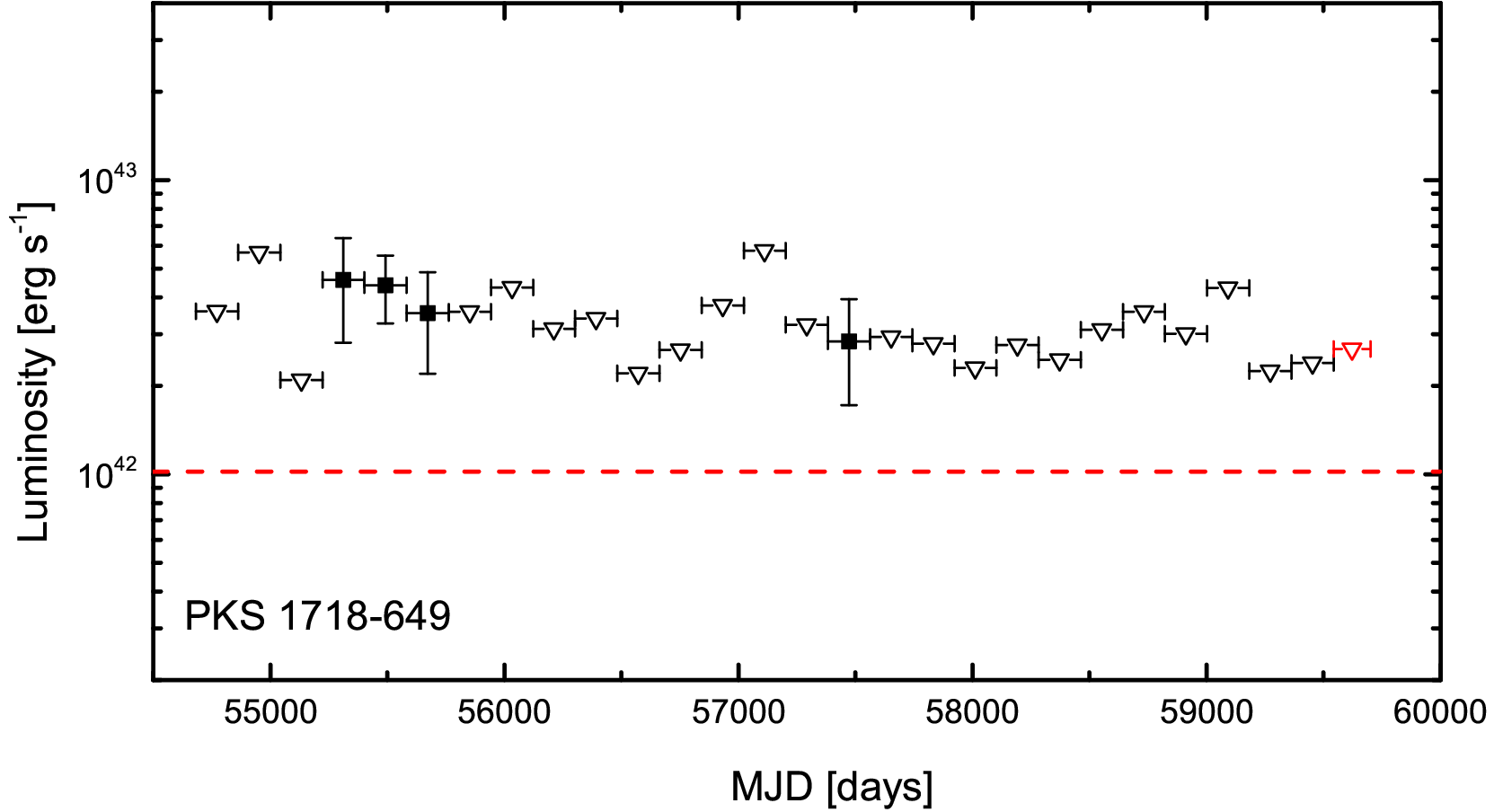}
   \includegraphics[angle=0,scale=0.27]{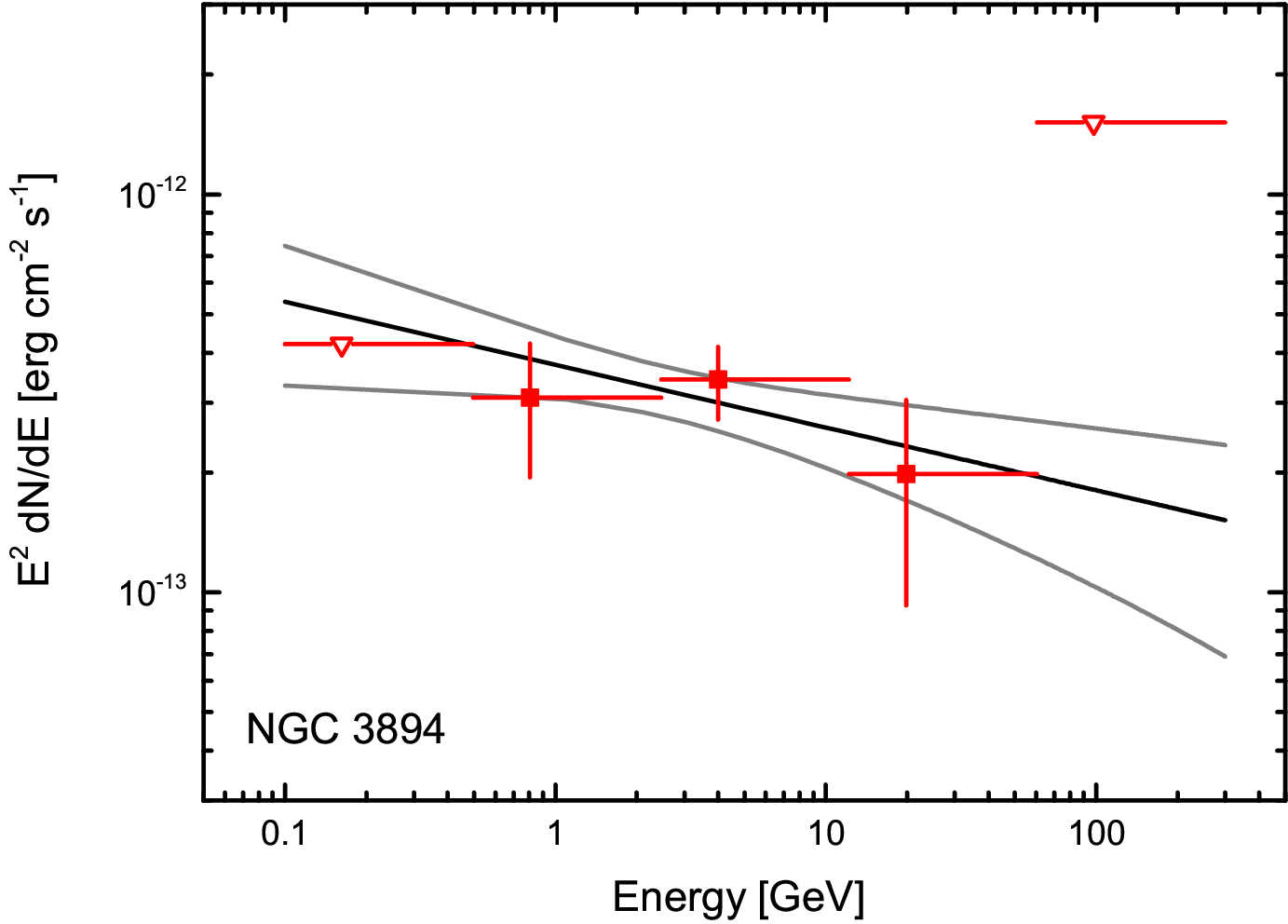}
   \includegraphics[angle=0,scale=0.3]{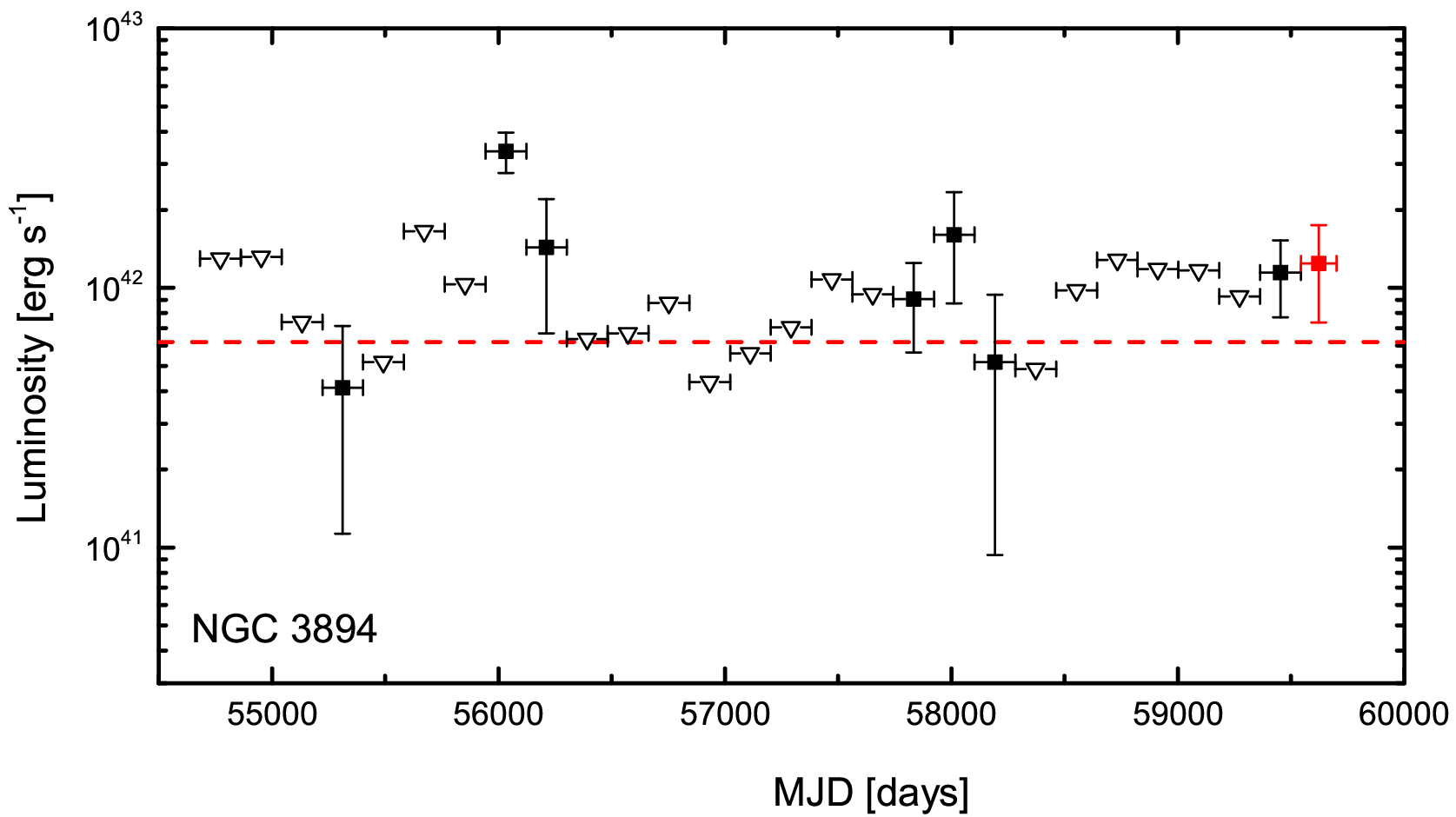}
   \includegraphics[angle=0,scale=0.27]{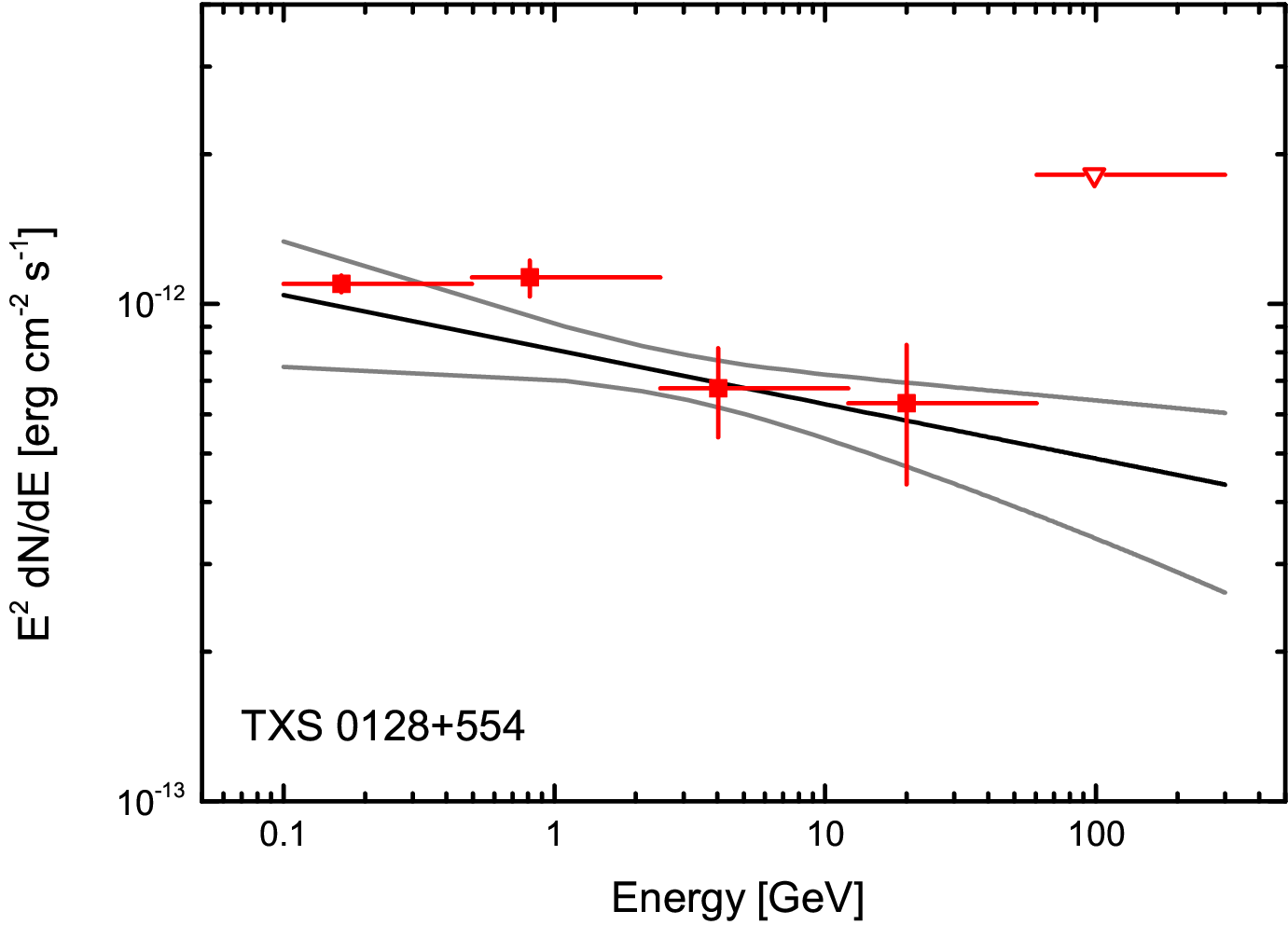}
   \includegraphics[angle=0,scale=0.3]{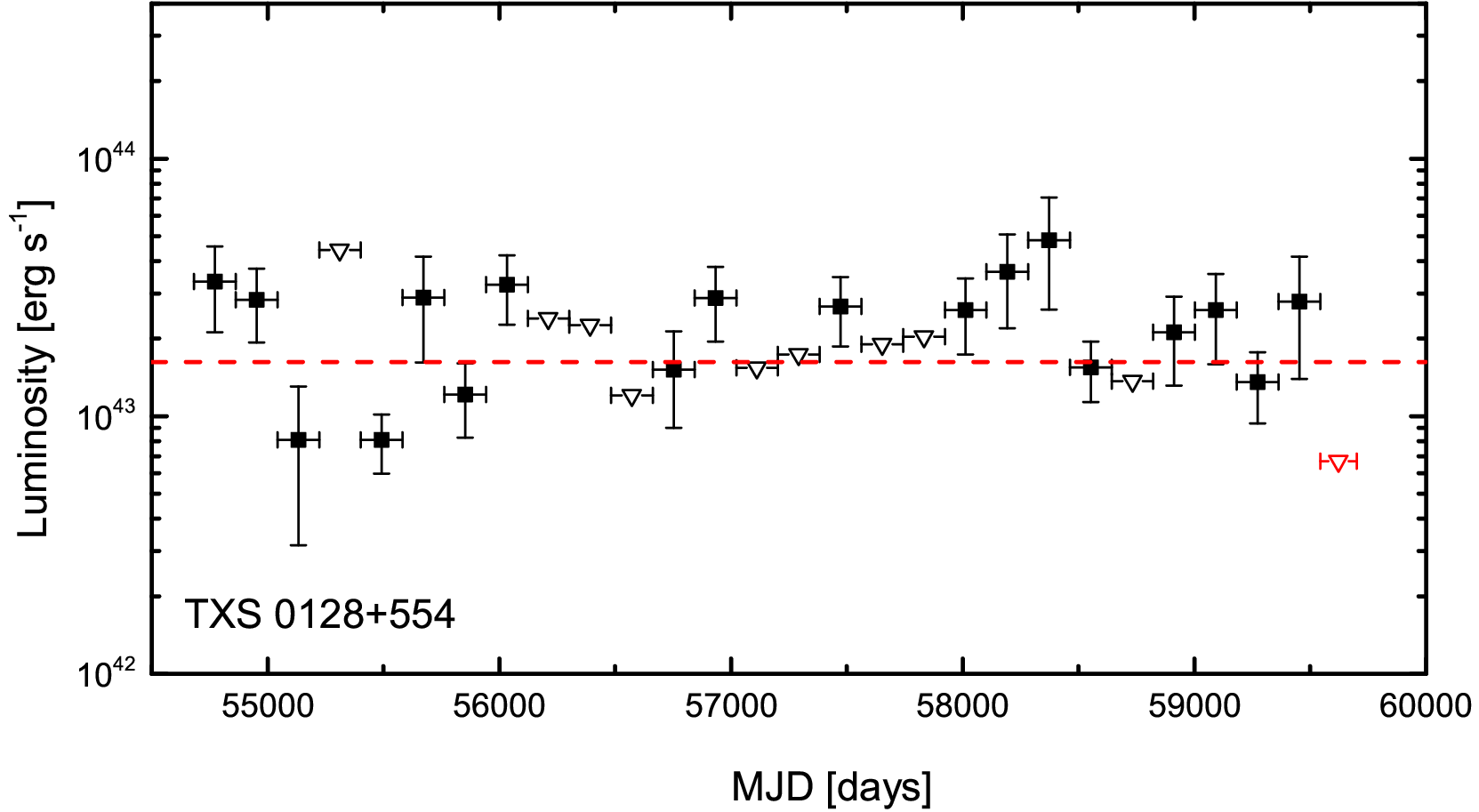}
\caption{Left column: the $\sim$14-yr Fermi/LAT average spectra along the power-law function fit results. Right column: the long-term Fermi/LAT light curves with time bins of 180 days. The horizontal red dash lines represent the $\sim$14 yr average $\gamma$-ray luminosity of sources, and the red data points in the light curves denote the last residual time interval, only 157 days. The opened inverted triangles indicate the TS value of that energy/time bin less than 9.}
\label{Spectra-LC}
\end{figure}

\emph{PKS 1718--649}. By analyzing the $\sim$14-yr Fermi/LAT observation data, a $\gamma$-ray detection with TS $\sim$ 37 is yielded for PKS 1718--649. As shown in Figure \ref{TSmap}, the radio position of PKS 1718--649 falls within the 68\% error circle of the best-fit position of 4FGL J1724.2--6501. Our result is coincident with the previous report that PKS 1718--649 is spatially associated with the $\gamma$-ray point source 3FGL J1728.0--6446 (\citealt{2016ApJ...821L..31M}), where 3FGL J1728.0--6446 is associated with 4FGL J1724.2--6501 in the 4FGL-DR3 (\citealt{2022ApJS..260...53A}). PKS 1718--649 has a steep spectrum at the GeV band with a photon spectral index of $\Gamma_{\gamma}=2.49\pm0.16$. Its $\sim$14-yr average luminosity at the GeV band is $(1.02\pm0.19)\times10^{42}$ erg s$^{-1}$. No obvious variability is observed with $\rm TS_{var}=35.6$, only four detection points are presented in the whole light curve, and the first three detection points are obtained with the first 7-yr observations. Using the 7-yr data set, \cite{2016ApJ...821L..31M} therefore found the $\gamma$-ray signal of PKS 1718--649 with the same TS value (TS $\sim$ 36) as us.

\emph{NGC 3894}. A significant $\gamma$-ray signal with TS = 96.4 is detected around NGC 3894. The radio position of NGC 3894 is located at the 68\% containment of the best-fit position of 4FGL J1149.0+5924, as displayed in Figure \ref{TSmap}. Its spectrum at the GeV band is flatter than that of PKS 1718--649, i.e., $\Gamma_{\gamma}=2.16\pm0.11$, which is slightly steeper than the derived value of $\Gamma_{\gamma}=2.01\pm0.10$ with the 10.8-yr observation data in \cite{2020A&A...635A.185P}. NGC 3894 has the lowest $\gamma$-ray luminosity among the three CSOs, being $(6.18\pm0.89) \times 10^{41}$ erg s$^{-1}$. The confidence level of the flux variation is less than 2$\sigma$, indicating no variability in its $\gamma$-ray emission too.

\emph{TXS 0128+554}. The analysis with the $\sim$14-yr Fermi/LAT observation data yields a significant $\gamma$-ray detection with TS=146.5 for TXS 0128+554. It has the highest $\gamma$-ray luminosity of $(1.62\pm0.17)\times 10^{43}$ erg s$^{-1}$ and the flattest spectrum with $\Gamma_{\gamma}=2.11\pm0.08$ among the three CSOs. Same as PKS 1718--649 and NGC 3894, no significant flux variation of $\gamma$-rays more than the 2$\sigma$ confidence level is observed for TXS 0128+554.

Note that \cite{2021MNRAS.507.4564P} investigated the $\gamma$-ray emission of some young radio sources by analysing the 11-yr Fermi/LAT data. They reported that TS$=36$ and $\Gamma_{\gamma}=2.60\pm0.14$ for PKS 1718--649, TS$=95$ and $\Gamma_{\gamma}=2.05\pm0.09$ for NGC 3894, TS$=178$ and $\Gamma_{\gamma}=2.20\pm0.07$ for TXS 0128+554, and found no evidence of significant $\gamma$-ray variability for the three CSOs. These results are consistent with our analysis results.

\subsection {The $L_{\gamma}-\Gamma_{\gamma}$ Plane} 

\begin{figure}
 \centering
   \includegraphics[angle=0,scale=0.45]{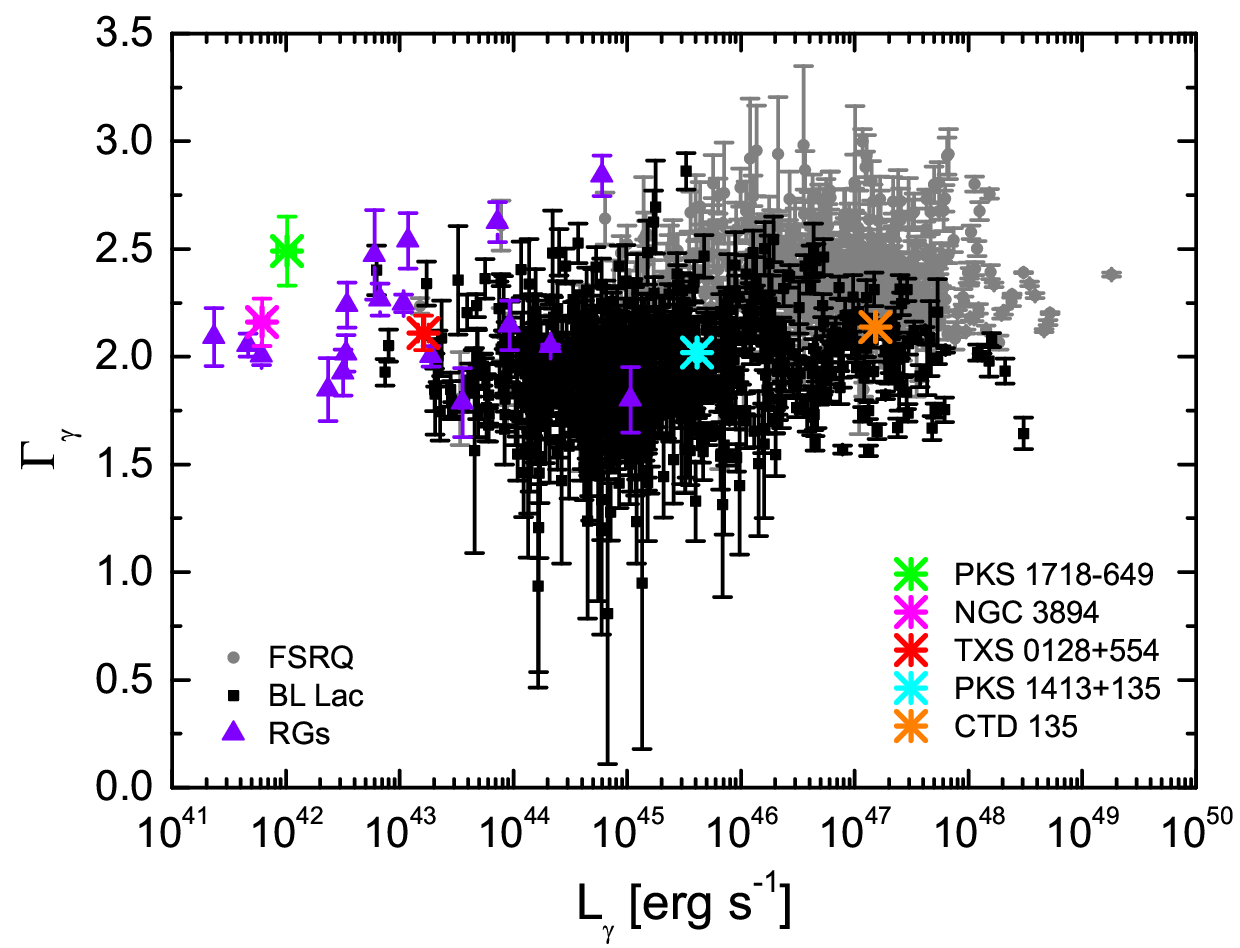}
\caption{The photon spectral index ($\Gamma_{\gamma}$) versus the $\gamma$-ray average luminosity ($L_{\gamma}$) in the LAT energy band. The data of blazars and RGs are taken from the 4FGL-DR3. The data of CTD 135 and PKS 1413+135 are from \cite{2021RAA....21..201G, 2022ApJ...939...78G}.}
\label{Gamma-L}
\end{figure}

As described above, the $\gamma$-ray average spectra of the three CSOs can be well fitted with a power-law function, none of them displays the significant flux variation in the GeV band, and all of them show the low $\gamma$-ray luminosity. These are dramatically different from the $\gamma$-ray emitting CSOs CTD 135 and PKS 1413+135, both of them show the Log-parabola spectra, violent variability, and high luminosity in the GeV band, similar to blazars (\citealt{2021RAA....21..201G, 2022ApJ...939...78G}). The diversity of the $\gamma$-ray emitting CSOs is also displayed in the $L_{\gamma}-\Gamma_{\gamma}$ plane, where $L_{\gamma}$ is the $\gamma$-ray luminosity in the GeV band, and the data of blazars and RGs are taken from the 4FGL-DR3 (\citealt{2022ApJS..260...53A}). As illustrated in Figure \ref{Gamma-L}, blazars together with CTD 135 and PKS 1413+135 occupy the high-luminosity area of the figure while RGs and three CSOs studied in this work are located at the low-luminosity region. It is widely known that blazars demonstrate frequent and violent variabilities, especially in the $\gamma$-ray band, favoring a compact emission region with significant Doppler boosting effect. Similarly, the high-luminosity $\gamma$-ray emission of CTD 135 and PKS 1413+135 is due to the episodic nuclear jet activities and is generated from a new-born aligned core-jet (\citealt{2021RAA....21..201G, 2022ApJ...939...78G}). In contrast, RGs normally have the low luminosity and weak variability, and they are thought to be the parent populations of blazars with large viewing angles and small Doppler factors (\citealt{1995PASP..107..803U}). Accordingly, it is still an open question whether the $\gamma$-rays of RGs originate from the jet with a large inclination angle or the large-scale jet extended regions. In the LAT band, Fornax A shows no significant variability while Cen A only displays some flux variation at a $\sim3.5\sigma$ confidence level in the low-energy band (\citealt{2017PhRvD..95f3018B, 2022ApJS..260...53A}). In particular, the Fermi/LAT observations definitively demonstrate the $\gamma$-rays from the large-scale lobes of Cen A and Fornax A, to account for more than 50\% and 86\% of their total $\gamma$-ray flux, respectively (\citealt{2010Sci...328..725A, 2016ApJ...826....1A}). Thereby, the $\gamma$-rays of Cen A and Fornax A are lobe-dominated. In view of no obvious $\gamma$-ray variability and the lobe-dominated radio emission for TXS 0128+554, PKS 1718--649, and NGC 3894, similar to Cen A and Fornax A, we speculate that their $\gamma$-rays stem from the extended mini-lobes.  

\section {Discussion} 

\subsection {Broadband SED Modeling and $\gamma$-ray Origin}

\begin{figure}
 \centering
   \includegraphics[angle=0,scale=0.21]{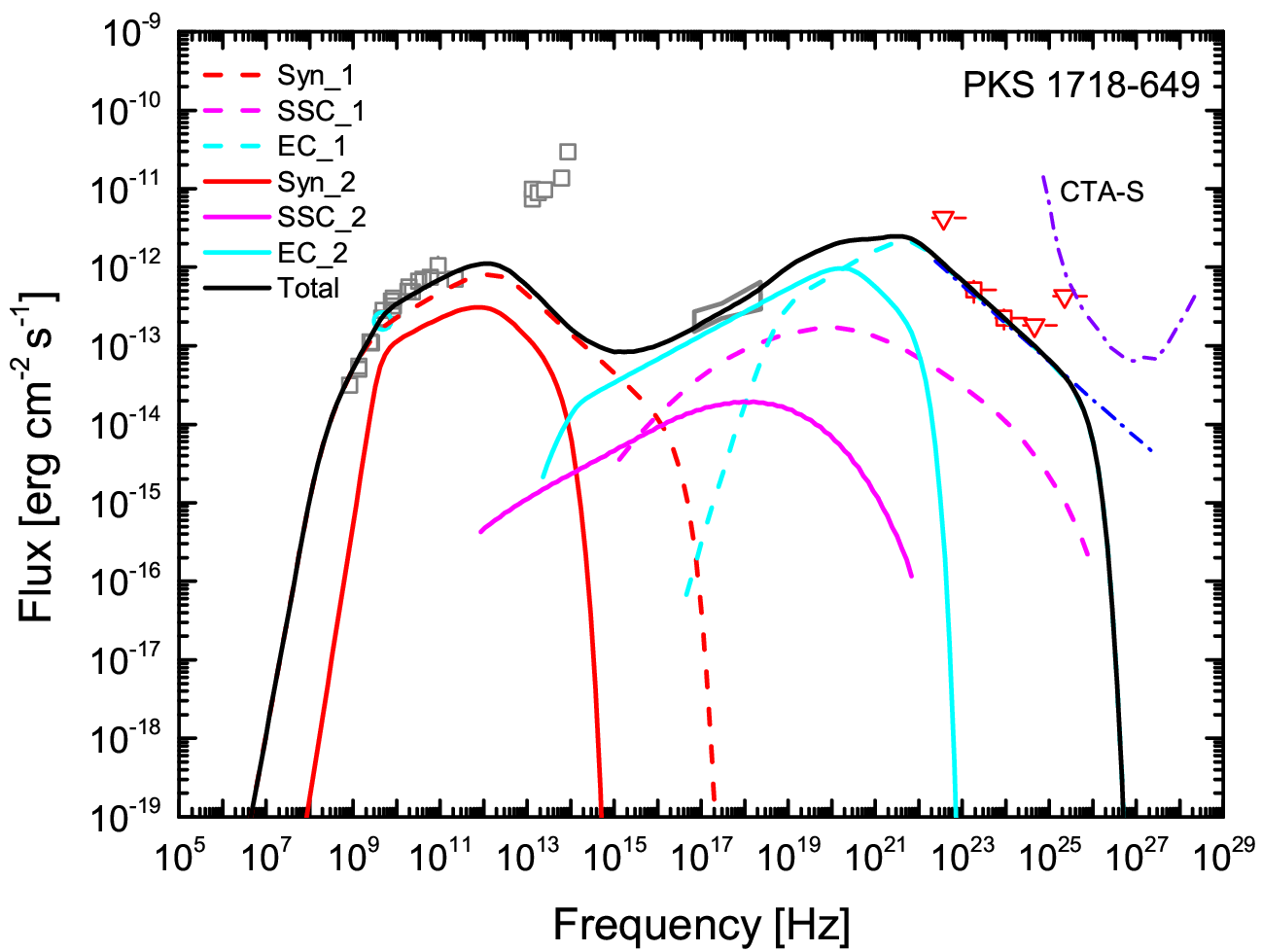}
   \includegraphics[angle=0,scale=0.21]{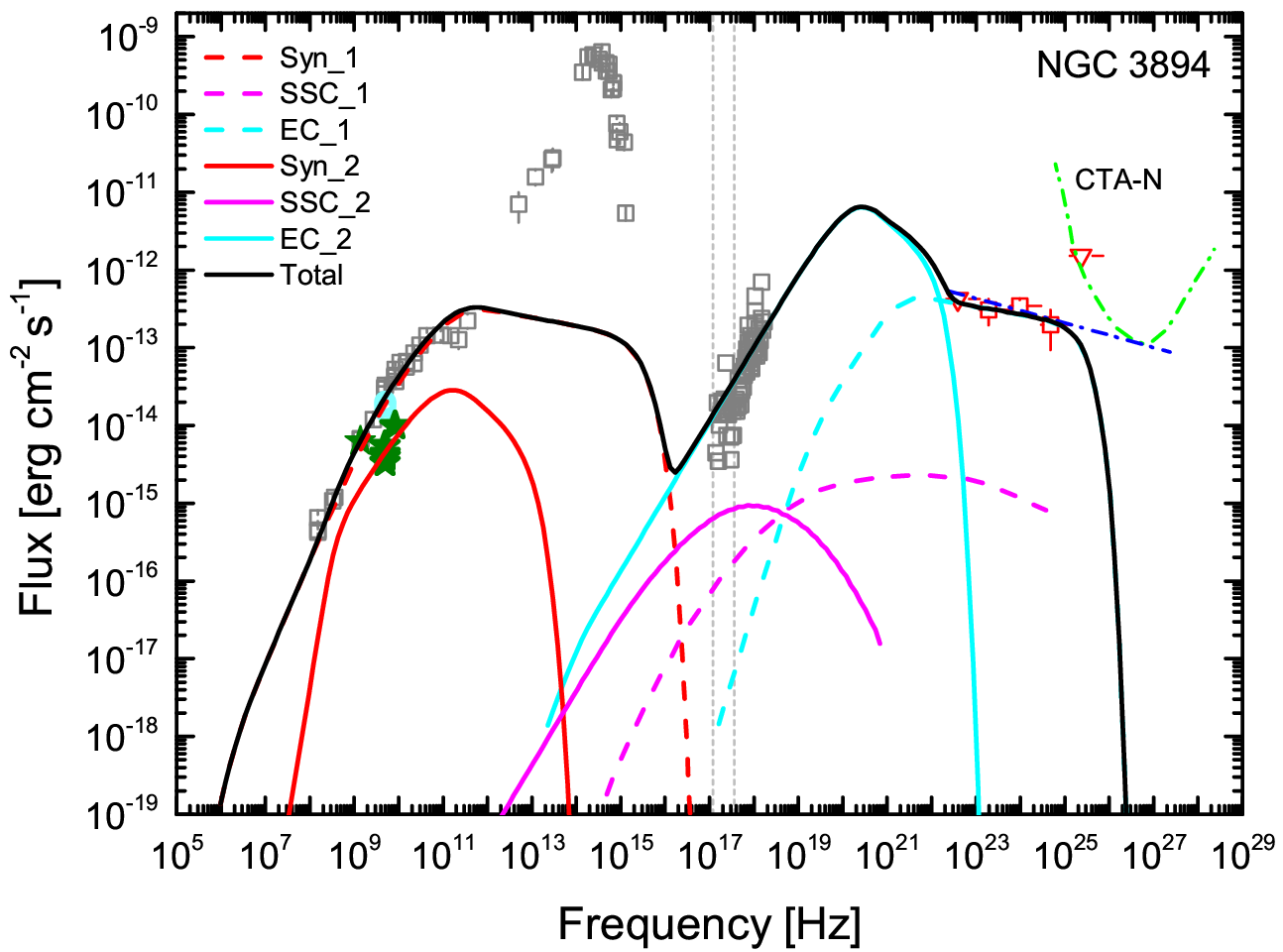}
   \includegraphics[angle=0,scale=0.21]{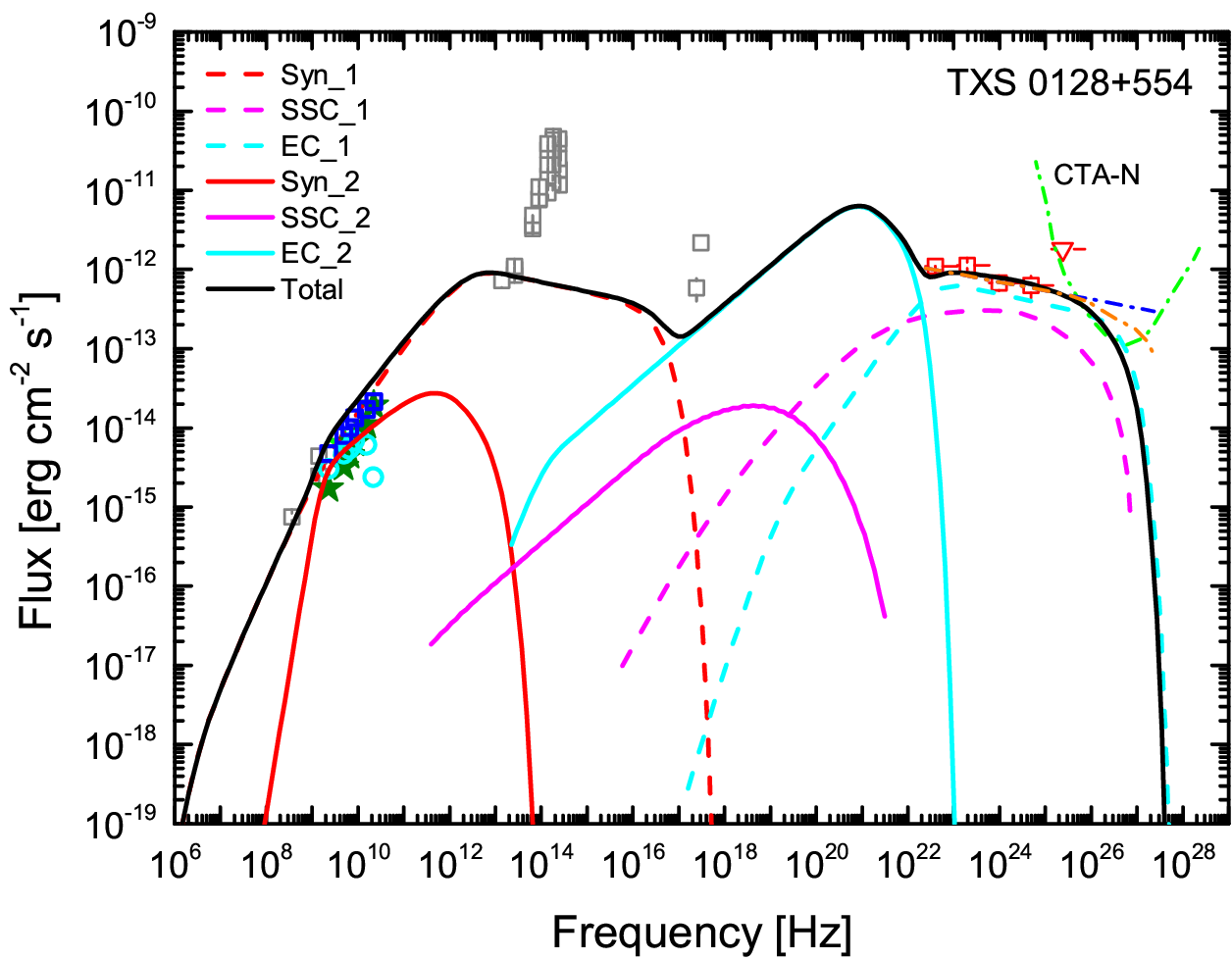}
\caption{Observed SEDs with the two-zone (`1' for extended region and `2' for core region) leptonic model fitting results. The data marked as opened grey squares are taken from the NED, ASDC, and literature (\citealt{2021ApJ...922...84B, 2022ApJ...941...52S}). The red squares and inverted triangles denote the Fermi/LAT average spectra obtained in this work. The olive stars (core) and the opened cyan circles (lobe) represent the flux of core and lobe, respectively, which are collected from the NED and literature (\citealt{1997AJ....113.2025T, 1998ApJ...498..619T, 2020ApJ...899..141L}). The opened blue squares indicate total flux taken from \cite{2020ApJ...899..141L}. The blue dash-dotted lines show the extrapolated VHE spectra from the Fermi/LAT spectra, while the orange dash-dotted line in TXS 0128+554 panel represents the extrapolating VHE spectrum considered the EBL absorption. The data of the soft X-ray band between two vertical lines for NGC 3894 are not considered during SED modeling.}
\label{SED}
\end{figure}

In order to further explore the $\gamma$-ray emission properties of the three CSOs, we collect the low-energy band data from literature and the NASA/IPAC Extragalactic Database (NED, \citealt{NASA/IPAC Extragalactic Database (NED). 2019}) as well as the ASI Science Data Center (ASDC) and construct their broadband spectral energy distributions (SEDs), as displayed in Figure \ref{SED}. Detailed description of data see the caption of Figure \ref{SED}. Considering the similar radiation properties of the three CSOs to RGs Cen A and Fornax A, the lobe-dominated emission in radio band and no significant variability in GeV band, we thus infer that the $\gamma$-rays of the three CSOs are also dominated by the extended region. Due to the broad second bump in the broadband SEDs of the three CSOs, we thus intend to represent their SEDs with the two-zone leptonic model, similar to that in other $\gamma$-ray emitting compact radio sources (e.g., \citealt{2020ApJ...899....2Z, 2022ApJ...927..221G, 2021RAA....21..201G, 2022ApJ...939...78G}). The synchrotron (syn), synchrotron-self-Compton (SSC), and external Compton (EC) scattering of the relativistic electrons in both the core and extended region are considered. The synchrotron-self-absorption (SSA) effect at the low-energy band, the Klein--Nishina effect at the high-energy band, and the absorption of high-energy $\gamma$-ray photons by extragalactic background light (EBL; \citealt{2008A&A...487..837F}) are also taken into account during the SED modeling.

\begin{figure}
 \centering
   \includegraphics[angle=0,scale=0.25]{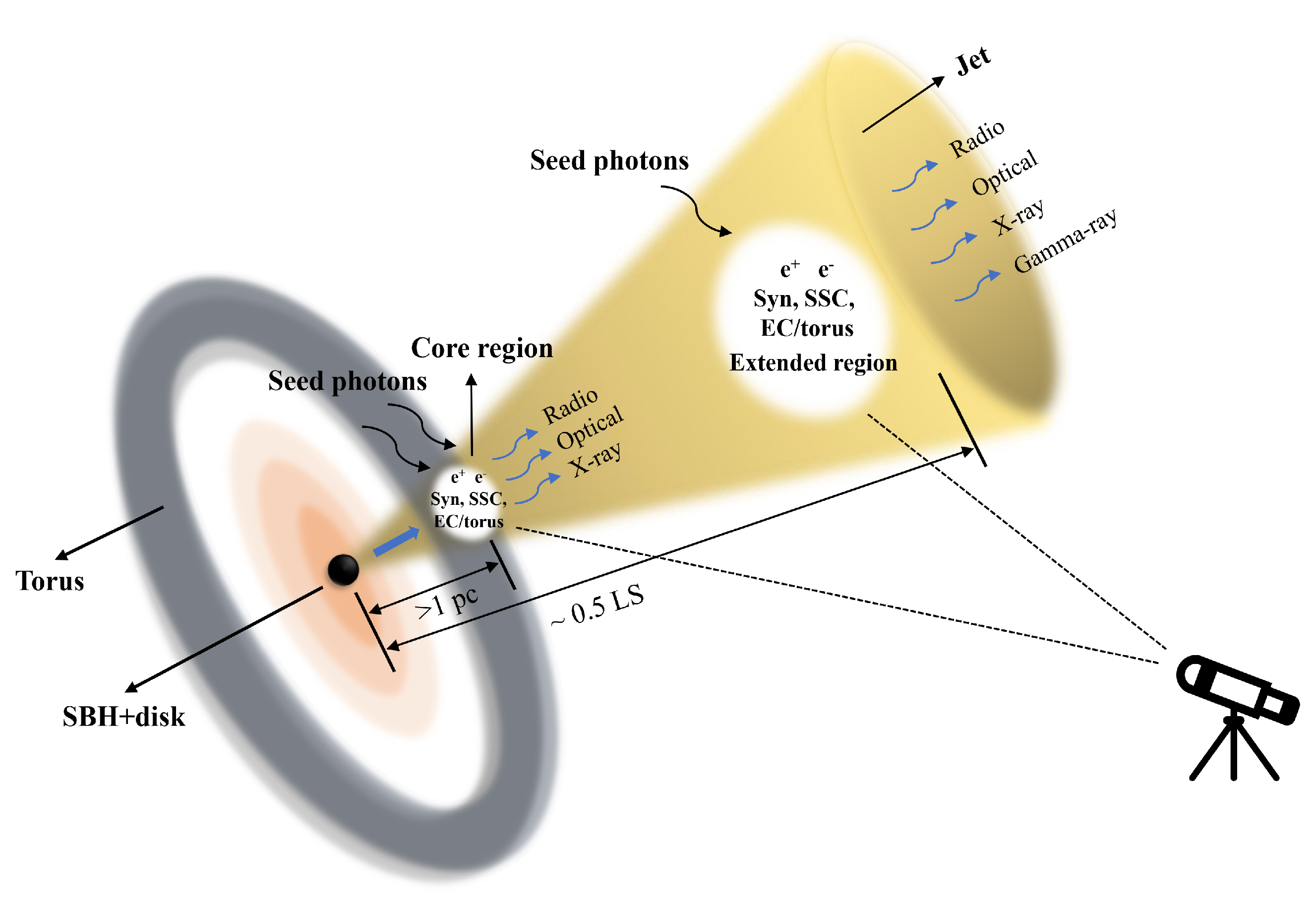}
\caption{A cartoon illustration of the two-zone leptonic radiation model. The distance between the core region and the black hole is conservatively estimated to be greater than 1 pc, with a rough approximation of $10*R$. The distance between the extended region and the black hole is approximately half of the total projection size of the radio morphology, where the term `LS' refers to the overall projection size of sources.}
\label{model}
\end{figure}

The electron distributions in both core and extended region are assumed as a broken power-law, which are characterized by a density parameter $N_0$, a break energy $\gamma_{\rm b}$, and indices $p_1$ and $p_2$ in the energy range of [$\gamma_{\min}, \gamma_{\max}$]. The radiation region is assumed as a sphere with radius $R$, magnetic field $B$, and Doppler factor $\delta$. Since the mildly relativistic motions have been observed and reported for NGC 3894 (\citealt{2020A&A...635A.185P}) and TXS 0128+554 (\citealt{2020ApJ...899..141L}), we take their relativistic effect into account for both core and extended region, i.e., $\delta=1.3$ and $\Gamma=1.04$ for NGC 3894 (\citealt{2020A&A...635A.185P}), $\delta=1.2$ and $\Gamma=1.06$ for TXS 0128+554 (corresponding to a viewing angle of $\theta=52\degr$, \citealt{2020ApJ...899..141L}), where $\Gamma$ is the bulk Lorenz factor. The jet is accelerated and collimated reaching up to sub-pc/pc scale (see \citealt{2017A&ARv..25....4B} for a review). After the acceleration phase, the jet assumes a conical shape and initiates deceleration as a logarithmic function of distance ($r$) from the black hole (see \citealt{2013MNRAS.429.1189P, 2022PhRvD.105b3005W}). We assume the same value of $\Gamma$ for both core and extended regions due to the total projection size of NGC 3894 and TXS 0128+554 being less than 20 pc. Assuming the same advance speed for the two-sided mini-lobes, an apparent advance velocity of only $\sim0.06~c$ is observed for PKS 1718--649 (\citealt{2019A&A...627A.148A}), we thus take $\Gamma=\delta=1$ for both core and extended region during SED modeling. For the core region, we take $R=\delta c \Delta t/(1+z)$ and $\Delta t=180$ days. Due to the absence of apparent variability in the $\gamma$-ray light curves, the time-bin value of the light curves is adopted, similar to other compact radio sources emitting $\gamma$-rays (\citealt{2020ApJ...899....2Z}). In this case, $R>0.1$ pc, the energy dissipation region should be located at $\sim10*R$ from the black hole and outside of the broad-line regions, and the photon field of torus provides the dominant seed photons of EC process (see \citealt{2009MNRAS.397..985G}). The overall projection size of the radio morphology of the three CSOs is all less than 20 pc. The seed photon field of EC process is still dominated by the photons from torus within this scale (\citealt{2009MNRAS.397..985G}, see also \citealt{2008ApJ...680..911S}). We thus only consider the external photon field of torus to calculate the EC process for both the core and extended region. The typical energy density of the torus photon field of $U_{\rm IR}=3\times10^{-4}$ erg cm$^{-3}$ (\citealt{2007ApJ...660..117C, 2014ApJS..215....5K}) is taken for the core region, whereas the values of $U_{\rm IR}$ for the extended regions are estimated according to Equation (21) in \cite{2008ApJ...680..911S} and are given in Table 4. For the core region, $\gamma _{\rm min}=1$ and $\gamma _{\rm max}=10^4$ are set. $p_1$ is roughly estimated by the spectral index of the X-ray observation data while $p_2=4$ is fixed. $\gamma_{\rm b}$, $N_0$, and $B$ are free parameters. For the extended region, $R$ is equal to the total projection size of the radio morphology, as listed in Table 2. $\gamma_{\rm min}$ is taken as 300 or 500 to match the observation data at radio band and $\gamma_{\rm max}$ is set as $\gamma_{\rm max}=100\gamma_{\rm b}$ (roughly constrained by the last upper-limit data point in the Fermi/LAT spectrum). $p_1$ and $p_2$ are constrained by the radio and $\gamma$-ray spectral indices, respectively. $\gamma_{\rm b}$, $N_0$, and $B$ are also free parameters. As shown in Figure \ref{SED}, the observational data at the near-IR-optical band are dominated by the starlight of host galaxy. The data of the soft X-ray band (0.5--1.5 keV) for NGC 3894 are not corrected the absorption (\citealt{2021ApJ...922...84B}). We thus do not take these data into account during the SED modeling. Due to the limited observation data, we cannot give a constraint on the fitting parameters, and we only obtain a set of model parameters that can make an acceptable fit. We also plot a cartoon illustration for the radiation model, as depicted in Figure \ref{model}.

The fitting results are presented in Figure \ref{SED} and the modeling parameters are given in Table 2. The broadband SEDs of the three CSOs can be well explained by the two-zone leptonic radiation model. Under this scenario, the $\gamma$-rays are produced by the IC process of the relativistic electrons in extended region and the X-rays are dominated by the radiation of core region. As listed in Table 2, the derived $B$ values of core regions are higher than that of extended regions while $\gamma_{\rm b}$ of core is smaller than that of extended region. We find that the magnetic field strengths between core region ($B_1$) and extended region ($B_2$) for the three CSOs basically satisfy the relation of $B_1/B_2 = r_2/r_1$, where $r_1$ and $r_2$ are the distances from the core and the extended regions to the black hole, respectively. This result is consistent with the VLBA observations for some AGN jets (\citealt{2009MNRAS.400...26O}). The derived $B$ values of extended regions are several mG, being in agreement with the typical values measured in CSOs (\citealt{2006A&A...450..959O, 2020ApJ...899..141L}). As listed in Table 2, $\gamma_{\max}\sim10^6$ is required to account for the $\gamma$-ray emission originating from these extended regions of CSOs. The detection of GeV $\gamma$-rays in the large-scale lobes of RGs Cen A and Fornax A (\citealt{2010Sci...328..725A, 2016ApJ...826....1A}) provides compelling evidence for the acceleration of ultrarelativistic electrons in the extended regions of AGN jets, particularly highlighted by the very high-energy detection along the jet of Cen A (\citealt{2020Natur.582..356H}). The leptonic radiation model employed to explain the observed $\gamma$-ray flux from these extended regions of AGN jets requires electron Lorentz factors on the order of $10^6$ (\citealt{2019MNRAS.490.1489P}), or even higher ($10^7-10^8$; \citealt{2020Natur.582..356H}).

It should be noted that we used a phenomenological electron spectrum, a broken power-law spectrum, to model the SEDs here. Although a power-law electron spectrum with an exponential cut-off would be more naturally motivated by particle acceleration theory, a broken power-law electron spectrum can better represent the symmetric peaks in the observed SEDs (e.g., \citealt{2009ApJ...696L.150A, 2009MNRAS.397..985G, 2012ApJ...752..157Z, 2012ApJ...748..119C, 2014A&A...563A..90A}). It is unclear whether the electron distribution has a low-energy cutoff with $\gamma_{\min}$>1 in theory; generally, it is fixed as $\gamma_{\min}=1$ (e.g., \citealt{2009MNRAS.397..985G, 2012ApJ...752..157Z}). Sometimes, $\gamma_{\min}$ can be roughly constrained with X-ray data (e.g., \citealt{2000ApJ...543..535T, 2014ApJ...788..104Z}), or a low-energy cutoff in the electron spectrum is required to explain the radio spectra of some large-scale jet substructures, with $\gamma_{\min}$ being several hundreds (\citealt{2000ApJ...530L..81H, 2001MNRAS.323L..17H, 2006ApJ...642L..33L, 2009ApJ...695..707G}). During SED modeling, $\gamma_{\max}$ normally is constrained by the last observation point in the $\gamma$-ray band. The theoretical upper limit of the maximum electron energy $\gamma_{\max}$ can be determined by $t_{\rm acc}$ = min$(t_{\rm esc}, t_{\rm cool})$, where $t_{\rm acc}$, $t_{\rm esc}$, $t_{\rm cool}$ are the timescales of acceleration, escape and cooling, respectively. We estimated the three theoretical timescales using the Equations from \cite{2022PhRvD.105b3005W}, and found that the derived values of $\gamma_{\max}$ from SED modeling are much smaller than the theoretical upper limit of $\gamma_{\max}$. In addition, $\gamma_{\rm b}$ derived from SED modeling is not exactly the same as the cooling break since $\gamma_{\rm b}$ results from a complex physical process, including the adiabatic losses, the particle escape, and the cooling (e.g., \citealt{2009MNRAS.397..985G}).

The energy densities of the non-thermal electrons ($U_{\rm e}$) and magnetic fields ($U_{B}$) of extended regions are also listed in Table 4. The ratio of $U_{\rm e}/U_{B}$ of the three CSOs is close to that of the lobes in Cen A and Fornax A, which are also derived by the SED modeling (\citealt{2019MNRAS.490.1489P}), and is also agreement with the values in the radio lobes of some RGs, which are obtained by the X-ray measurements (e.g., \citealt{2005ApJ...622..797K, 2005ApJ...626..733C, 2011ApJ...727...82I}). 

Based on the SED fitting parameters, we estimate the jet powers ($P_{\rm jet}^{e^{\pm}}$) of core and extended region in the case of the electron–positron pair jets. The powers of the non-thermal electrons ($P_{\rm e}$) and magnetic fields ($P_{B}$) are calculated with $P_{\rm i} = \pi R^{2}\Gamma^{2}cU_{\rm i}$, where $U_{\rm i}$ can be $U_{\rm e}$ or $U_{B}$. The radiation power of jets is estimated by $P_{\rm r} = \pi R^{2}\Gamma^{2}cU_{\rm r} = L_{\rm bol}\Gamma^{2}/4\delta^4$, where $L_{\rm bol}$ is the total luminosity of non-thermal radiation from a radiation region. $P_{\rm jet}^{e^{\pm}}$ is the sum of $P_{\rm e}$, $P_{B}$, and $P_{\rm r}$. As given in Table 3 and 4, $P_{\rm jet}^{e^{\pm}}$ of core is within $3\times10^{42}-10^{44}$ erg s$^{-1}$ while $P_{\rm jet}^{e^{\pm}}$ of extended regions are in the range of $\sim10^{43}-10^{44}$ erg s$^{-1}$. It is found that the jet powers of core and extended region for the three CSOs are roughly of the same order, and similar results have been reported for a $\gamma$-ray-emitting RG 3C 120 (\citealt{2017A&A...608A..37Z}) and a blazar 4C +49.22 (\citealt{2018ApJ...865..100Z}). We also estimate the kinetic power ($P_{\rm kin}$) of the three CSOs with the radio luminosity at 1.4 GHz of extended region using Equation (1) in \cite{2010ApJ...720.1066C} and obtain $1.4\times10^{44}$ erg s$^{-1}$, $3.0\times10^{42}$ erg s$^{-1}$ and $7.3\times10^{43}$ erg s$^{-1}$ for PKS 1718--649, NGC 3894 and TXS 0128+554, respectively. $P_{\rm kin}$ is also roughly of the same order as $P_{\rm jet}^{e^{\pm}}$ of core and extended region.

\subsection {Possible VHE Emission}

Recently, CSO PKS 1413+135 is detected at the very high energy (VHE) band by the MAGIC telescope (\citealt{2022ATel15161....1B, 2022ApJ...939...78G}). Although the $\gamma$-rays of PKS 1413+135 should be dominated by an aligned jet emission and may be totally different from the three lobe-dominated CSOs, we still wondering whether the three CSOs could be detected at the VHE band. In particular, the H.E.S.S. collaboration reported that Cen A has the VHE emission along its large-scale jet, providing the direct evidence for the presence of high-energy electrons in the extended regions of AGN jets (\citealt{2020Natur.582..356H}).  

Extrapolating the Fermi/LAT $\gamma$-ray spectrum to the VHE band is a common method to estimate the intrinsic VHE spectrum of sources (\citealt{2021MNRAS.508.6128P, 2021ApJ...916...93Z, 2022MNRAS.515.4505M}). As described in Section 3.2, the spectra observed by the Fermi/LAT for the three CSOs are well represented by a power-law function, and no evidence of variability has been found in their long-term $\gamma$-ray light curves. We therefore directly extend the power-law spectra at the GeV band to the VHE band for the three CSOs, as illustrated in Figure \ref{SED}. The sensitivity curves of Cherenkov Telescope Array south and north (CTA-S and CTA-N, 50 hr) are also presented in Figure \ref{SED}, where the sensitivity curves are obtained from the CTA webpage\footnote{https://www.cta-observatory.org/science/cta-performance/}. It can be found that the extrapolated flux at the VHE band of PKS 1718--649 and NGC 3894 is below the sensitivity curve of CTA. Especially for PKS 1718--649, together with its the steep spectrum at the GeV band, it would be difficult to be detected by the CTA in future. The extrapolated intrinsic VHE spectrum of TXS 0128+554 clearly surpasses the sensitivity curve of the CTA; however, when considering the EBL absorption, it only marginally intersects with the CTA's sensitivity curve, as depicted in Figure \ref{SED}. Nevertheless, theoretical suggestions have proposed episodic jet activity for compact radio sources (e.g., \citealt{1997ApJ...487L.135R, 2009ApJ...698..840C}), and the episodic nuclear jet activity has been reported for TXS 0128+554 (\citealt{2020ApJ...899..141L}). The reactivation of TXS 0128+554 could position it as a potential candidate for VHE emission, making it detectable by the CTA.

\section{Summary}

We completely analyzed the $\sim$14 yr Fermi/LAT observation data of three radio-lobe-dominated CSOs (PKS 1718--649, NGC 3894, TXS 0128+554) to explore their $\gamma$-ray emission properties. They are all characterised by a low $\gamma$-ray luminosity ($10^{41}-10^{43}$ erg s$^{-1}$), their average spectra in the 0.1--300 GeV can be well fitted by a power-law function, and no significant variability is observed in their long-term $\gamma$-ray light curves. These $\gamma$-ray features are prominently different from that of the $\gamma$-ray emitting CSOs PKS 1413+135 and CTD 135. In the $L_{\gamma}-\Gamma_{\gamma}$ plane, the three CSOs are located at the region occupied by RGs, also distinctly different from PKS 1413+135 and CTD 135, which are similar to blazars. Considering the similar radiation properties between the three CSOs and the $\gamma$-ray emitting RGs, specifically similar to RGs Cen A and Fornax A, together with no significant variability in the $\gamma$-ray band and no superluminal motion at the radio band, we speculated that their $\gamma$-rays stem from the lobe-dominated radiation, different from CTD 135 and PKS 1413+135 (\citealt{2021RAA....21..201G, 2022ApJ...939...78G}). Their broadband SEDs can be represented by the two-zone leptonic model and $\gamma$-rays are produced by the IC processes of the relativistic electrons in extended regions. The $\gamma$-rays are produced by the IC process of the relativistic electrons in extended region and the X-rays are dominated by the radiation of core region. This radiation model can be checked by studying whether the flux variations in X-ray and $\gamma$-ray bands are correlated. On the basis of the derived fitting parameters, we calculated the jet powers of core and extended region by assuming the electron–positron pair jet and found that the jet powers of core and extended region for the three CSOs are roughly of the same order. And, TXS 0128+554 may be a VHE emission candidate and be detected by CTA in future.

\begin{acknowledgements}
We thank the kind permission for the usage of the RFC data from the RFC Collaboration. This work is supported by the National Natural Science Foundation of China (grants 12022305, 11973050, 12203022).
\end{acknowledgements}

\clearpage

\begin{table}
\centering
\caption{Fermi/LAT Analysis Results}
\begin{threeparttable}
\begin{tabular}{lccccccc}
\hline\hline 
Source Name & $R.A.$\tnote{a} & $Decl.$\tnote{a} & TS & $L_{0.1-300~\rm GeV}$ & $\Gamma_{\rm \gamma}$ & TS$_{\rm var}$ & Association\\
& [deg] & [deg] & & [erg s$^{-1}$] & & & \\
\hline
PKS 1718--649&261.033&-65.0277&37.1&(1.02$\pm$0.19)E42&2.49$\pm$0.16&35.6 ($\sim1.3\sigma$)&4FGL J1724.2--6501\\
NGC 3894&177.254&59.4158&96.4&(6.18$\pm$0.89)E41&2.16$\pm$0.11&37.3 ($\sim1.5\sigma$)&4FGL J1149.0+5924\\
TXS 0128+554&22.8277&55.7705&146.5&(1.62$\pm$0.17)E43&2.11$\pm$0.08&34.0 ($\sim 1.1\sigma$)&4FGL J0131.2+5547\\
\hline
\end{tabular}
\begin{tablenotes}
    \footnotesize
    \item [a] The best-fit position of the $\sim$14 yr Fermi/LAT observation data obtained by the \emph{gtfindsrc} tool.
    \end{tablenotes}
\end{threeparttable} 
\end{table}

\begin{sidewaystable}[h]
\centering
\caption{SED Fitting Parameters}
\begin{tabular}{lcccccccccccccccccccccc}
\hline\hline 
 & \multicolumn{10}{c}{Compact Core} & &\multicolumn{10}{c}{Extended Region}
\\ [0.5ex]
\cline{2-11} \cline{13-22}
Source & $R$ & $B$ & $\delta$ & $\Gamma$ & $\gamma_{\min}$ & $\gamma_{\rm b}$ & $\gamma_{\max}$ & $N_{0}$ & $p_1$ & $p_2$ & & $R$ & $B$ & $\delta$ & $\Gamma$ & $\gamma_{\min}$ & $\gamma_{b}$ & $\gamma_{\max}$ & $N_{0}$ & $p_1$ & $p_2$ \\
& [cm] & [mG] & & & & & & [cm$^{-3}$] & & & & [pc] & [mG] & & & & & & [cm$^{-3}$] & & \\
\hline 
PKS 1718--649&4.60E17&80&1&1&1&1706&1.0E4&1321&2.40&4.00&&2.5&4.2&1&1&300&9.5E3&9.5E5&50&2.33&4.00\\
NGC 3894&6.00E17&10&1.3&1.04&1&1450&1.0E4&6.26E-2&1.08&4.00&&10.9&1.7&1.3&1.04&500&5.0E3&5.0E5&5.5E-6&1.02&3.20\\
TXS 0128+554&5.40E17&10&1.2&1.06&1&3056&1.0E4&477&2.00&4.00&&17.9&1.0&1.2&1.06&500&2.5E4&2.5E6&7.5E-5&1.25&3.25\\
\hline 
\end{tabular}
\end{sidewaystable}

\begin{table}
\centering
\caption{Derived Parameters of the Core Region}
\begin{tabular}{lccccccc}
\hline\hline 
Source & $U_{B}$ & $U_{\rm e}$ & $U_{\rm e}$/$U_{B}$ & $P_B$ & $P_{\rm e}$ & $P_{\rm r}$ & $P_{\rm jet}^{e^{\pm}}$ \\
& [erg cm$^{-3}$] & [erg cm$^{-3}$] & & [erg s$^{-1}$] & [erg s$^{-1}$] & [erg s$^{-1}$] & [erg s$^{-1}$] \\
\hline
PKS 1718--649&2.55E-4&2.59E-3&10.2&5.07E42&5.17E43&8.72E41&5.76E43\\
NGC 3894&3.98E-6&6.54E-5&16.4&1.46E41&2.40E42&6.15E41&3.16E42\\
TXS 0128+554&3.98E-6&3.31E-3&832.1&1.23E41&1.02E44&1.18E43&1.14E44\\
\hline
\end{tabular}
\end{table}

\begin{table}
\centering
\caption{Derived Parameters of the Extended Region}
\begin{tabular}{lccccccccc}
\hline\hline 
Source& $\nu_{\rm IR}$ & $U_{\rm IR}$ & $U_{B}$ & $U_{\rm e}$ & $U_{\rm e}$/$U_{B}$ & $P_B$ & $P_{\rm e}$ & $P_{\rm r}$ & $P_{\rm jet}^{e^{\pm}}$\\
& [Hz] & [erg cm$^{-3}$] & [erg cm$^{-3}$] & [erg cm$^{-3}$] & & [erg s$^{-1}$] & [erg s$^{-1}$] & [erg s$^{-1}$] & [erg s$^{-1}$] \\
\hline
PKS 1718--649&2.54E13&6.99E-7&7.02E-7&1.38E-5&19.7&3.93E42&7.76E43&2.11E42&8.36E43\\
NGC 3894&2.97E13&5.83E-8&1.15E-7&3.31E-8&0.3&1.33E43&3.82E42&1.44E41&1.72E43\\
TXS 0128+554&2.59E13&9.39E-9&3.98E-8&2.51E-7&6.3&1.28E43&8.12E43&6.07E42&1.00E44\\
\hline
\end{tabular}
\end{table}

\end{document}